\documentclass[12pt]{article}
\usepackage{epsf}\usepackage{amssymb} \usepackage{cite}
\usepackage{amsthm}
\usepackage{amscd}
\usepackage{amsfonts}
\usepackage{amsmath}
\newcommand{\C}{\mathbb C}

\newcommand{\Z}{\mathbb Z}

\newcommand{\be}{\begin{equation}}
\newcommand{\ee}{\end{equation}}
\newcommand{\bea}{\begin{eqnarray}}
\newcommand{\eea}{\end{eqnarray}}

\usepackage{hyperref}
\def\stackreb#1#2{\ \mathrel{\mathop{#1}\limits_{#2}}}
\renewcommand{\imath}{\mathrm{i}}

\begin{document}
\thispagestyle{empty}
\def\thefootnote{\fnsymbol{footnote}}
\begin{center}\Large \bf
Regge symmetry of 6j-symbols of the Lorentz group \\
\end{center}

\vskip 0.2cm

\begin{center}
Elena Apresyan$^{1}$\footnote{elena-apresyan@mail.ru} and
Gor Sarkissian$^{1,2}$
\footnote{gor.sargsyan@yerphi.am}
\end{center}
\begin{center}
$^1$Alikhanian National Laboratory,\\
Alikhanian Br. 2, 0036\, Yerevan, Armenia
\end{center}
\begin{center}
$^2$ Bogoliubov Laboratory of Theoretical Physics, JINR,\\
Dubna, Moscow region, 141980 Russia\\
\end{center}

\vskip 1.5em

\begin{abstract} \noindent
In this paper we derive new symmetry and new expression for $6j$-symbols of the unitary principal series
representations of the $SL(2,\mathbb{C})$ group.
This allowed us to derive for them the analogue of the Regge symmetry.
\end{abstract}

\newpage

\section{Introduction}

 T. Regge  has found 65 years ago \cite{Regge:1959ze}, that the $su(2)$ $6j$-symbols
$\big\{\,{}^{j_1}_{j_3}\,{}^{j_2}_{j_4}\,\big|\,{}^{j_{12}}_{j_{34}}\big\}$
besides obvious invariance under any permutations of the columns and  simultaneous interchange of upper and lower arguments in any two columns, additionally possess less trivial discrete symmetry
\be\nonumber
\big\{\,{}^{j_1}_{j_3}\,{}^{j_2}_{j_4}\,\big|\,{}^{j_{12}}_{j_{23}}\big\}=
\big\{\,{}^{S-j_1}_{S-j_3}\,{}^{S-j_2}_{S-j_4}\,\big|\,{}^{j_{12}}_{j_{23}}\big\}\,,
\ee
where $S={1\over 2}(j_1+j_2+j_3+j_4)$.
In other groundbreaking work \cite{pr}, Ponzano and Regge expanding on work of Wigner \cite{wiga}, conjectured remarkable asymptotic formula relating the value of the $6j$-symbols in the limit of the large spins to
the volume of the associated Euclidean tetrahedron. This conjecture was proved in \cite{Roberts:1998zka}.
Combining these two facts one arrives to the invariance
of the tetrahedron volume under the Regge's transformation. It is worth to emphasize that this property of the tetrahedron volume
was unknown before the
Regge's works.  Geometric proof can be found in \cite{ai}.

It turns out that the Regge symmetry persists when we consider the $6j$-symbols  of the various
generalization of $su(2)$ algebra.
First it was checked for the $6j$-symbols  of finite-dimensional representations of quantum algebra ${\rm U}_{\rm q}({\rm su}(2))$ \cite{Kirillov:1991ec,klimyk}.
Recently it was established in  \cite{Apresyan:2022erh}, that $6j$-symbols for the unitary principal series representations of the Faddeev modular double constructed
 in \cite{Ponsot:2000mt}, enjoy the Regge symmetry as well.

In this paper we study generalization of the Regge transformation for the $6j$-symbols of unitary principal series representation of $SL(2,\mathbb{C})$ group constructed
in \cite{Ismag2,Derkachev:2019bcl}.

Symmetries of the $SL(2,\mathbb{C})$ $6j$-symbols were studied
in papers \cite{Derkachov:2021thp,Sarkissian:2020ipg}.
Here we derive one more symmetry which in fact is composition of that symmetries. But we prefer to derive it here by a limiting
procedure from the corresponding symmetry property of the above mentioned $6j$-symbols of the Faddeev modular double, found in  \cite{Teschner:2012em},
and re-derived  and presented in \cite{Apresyan:2022erh} in much more simple and convenient for the application way.
Using this symmetry property we derive new expression for the $SL(2,\mathbb{C})$ $6j$-symbols
which allows us to deduce the corresponding Regge symmetry.

\section{6j-symbols for Lorentz group}
The unitary principal series
representations of the Lorentz group $SL(2,\mathbb{C})$ are labelled by pairs $(\sigma,N)$ where $\sigma\in \mathbb{R}$ and $N\in \mathbb{Z}$.
The group element
\be\nonumber
g=\left(\begin{array}{cc}
\alpha&\beta\\
\gamma&\delta \end{array}\right), \quad\alpha\delta-\beta\gamma=1
\ee
is represented by operator $T_a(g)$ acting on the space of the square-integrable single-valued functions $\Phi(z,\bar{z})$ on
$\mathbb{C}$ in the following way \cite{GGV}:
\be\nonumber
[T_a(g)\Phi](z,\bar{z})=(\beta z +\delta)^{a-1}(\bar{\beta} \bar{z} +\bar{\delta})^{\bar{a}-1}
\Phi\left({\alpha z+\gamma\over \beta z +\delta},{\bar{\alpha} \bar{z}+\bar{\gamma}\over \bar{\beta}\bar{z}  +\bar{\delta}}\right)\, .
\ee
For the unitary principal series representations one has
\be\nonumber
a={N\over 2}+i\sigma,\quad \bar{a}=-{N\over 2}+i\sigma,\quad N\in \mathbb{Z},\quad \sigma \in \mathbb{R}\, .
\ee
To write down the 6j-symbols for the $SL(2,\mathbb{C})$ group we need the complex gamma function \cite{GGV}:
\begin{equation}
{\bf \Gamma}(x,n)
=\frac{\Gamma(\frac{n+\textup{i}x}{2})}{\Gamma(1+\frac{n-\textup{i}x}{2})},
\label{Cgamma}\end{equation}
where $x\in \C $ and $n\in\Z$.
Define also so called complex hypergeometric function \cite{Derkachov:2021thp}:
\be\label{jdiscret}
\mathcal{J}_{cr}(\underline{s},\underline{n};\underline{t},\underline{m})
=\frac{1}{4\pi}\sum_{N\in\Z}\int_{-\infty}^{\infty}\prod_{a=1}^4{\bf \Gamma}(s_a-y,n_a-N)
{\bf \Gamma}(t_a+y,m_a+N)dy,
\ee
where the parameters
$n_a$, $m_a$, $s_a$ and $t_a$ satisfy the balancing condition:
\be\label{baldis}
\sum_{a=1}^4( n_a+m_a)=0,\qquad \sum_{a=1}^4( s_a+t_a)=-4\textup{i}.
\ee
Using these definitions the 6j-symbols for $SL(2,\mathbb{C})$ group can be written as \cite{Ismag2,Derkachev:2019bcl,Derkachov:2021thp}:
\bea \nonumber && \makebox[-1em]{}
\big\{\,{}^{\sigma_1,N_1}_{\sigma_3,N_3}\,{}^{\sigma_2,N_2}_{{\sigma}_4,N_4}\,|\,{}^{\rho_1,M_1}_{\rho_2,M_2}\big\}
={\pi^2\over 4}{{\bf \Gamma}\left(\sigma_1-\sigma_2+\rho_2-\textup{i}, A_1\right)
{\bf \Gamma}\left(\sigma_2-\sigma_3+\rho_1-\textup{i}, A_2\right)\over {\bf \Gamma}\left(-\sigma_3-\sigma_4+\rho_2-\textup{i}, A_3\right)
{\bf \Gamma}\left(\sigma_1+\sigma_4+\rho_1-\textup{i}, A_4\right)}
\\ && \makebox[-2em]{} \times
(-1)^{M_2-N_2+N_4}{\mathcal J}_{cr}(\underline{R},\underline{S};
\underline{U},\underline{T})\, ,
\label{idsgamma}\eea
where the function ${\mathcal J}_{cr}(\underline{R},\underline{S};
\underline{U},\underline{T})$ is defined by formula \eqref{jdiscret},
and
\begin{equation}\label{ra}
\begin{aligned}
&R_1=-\sigma_1+\sigma_2-\rho_2-\textup{i},\\
&R_2=\sigma_1+\sigma_2-\rho_2-\textup{i},\\
&R_3=-\sigma_3-\sigma_4-\rho_2-\textup{i},\\
&R_4=\sigma_3-\sigma_4-\rho_2-\textup{i},
\end{aligned}\quad
\begin{aligned}
&U_1=-\rho_1-\sigma_2+\sigma_4+\rho_2,\\
&U_2=\rho_1-\sigma_2+\sigma_4+\rho_2,\\
&U_3=0,\\
&U_4=2\rho_2,
\end{aligned}\quad
\begin{aligned}
&S_1=(-N_1+N_2-M_2)/2,\\
&S_2=(N_1+N_2-M_2)/2,\\
&S_3=-(N_3+N_4+M_2)/2,\\
&S_4=(N_3-N_4-M_2)/2,
\end{aligned}
\end{equation}
\begin{equation}\label{ta}
T_1=(-M_1-N_2+N_4+M_2)/2, \quad T_2=(M_1-N_2+N_4+M_2)/2, \quad T_3=0, \quad T_4=M_2,
\end{equation}
and
$$
A_1={N_1-N_2+M_2\over 2},\qquad A_2={N_2-N_3+M_1\over 2},
$$
$$ A_3={-N_3-N_4+M_2\over 2},\qquad A_4= {N_1+N_4+M_1\over 2}.
$$
Note that the parameters $R_a,U_a,S_a,T_a$ as they are defined in  \eqref{ra} and \eqref{ta} indeed satisfy the relation \eqref{baldis}:
\be\label{baldic}
\sum_{a=1}^4(R_a+U_a)=-4\textup{i}\quad{\rm and}\quad \sum_{a=1}^4(S_a+T_a)=0\, .
\ee
The discrete parameters $N_k$, $k=1,2,3,4$ and $M_1$, $M_2$ are constrained \cite{Naimark} to take only the values bringing
to integer values of $S_k$, $T_k$, $A_k$, k=1,2,3,4. In fact it is easy to see that it is enough to guarantee
that $A_k$ are integer and the rest of parameters will be integer automatically.

Define the following hyperbolic hypergeometric function \cite{Apresyan:2022erh,Ponsot:2000mt,rusik}
\be\label{hyj}
J_h(\underline{\mu},\underline{\nu})=\int_{-i\infty}^{i\infty}\prod_{a=1}^4\gamma^{(2)}(\mu_a-z;\omega_1,\omega_2)\gamma^{(2)}(\nu_a+z;\omega_1,\omega_2)
{dz\over i\sqrt{\omega_1\omega_2}}\, ,
\ee
where the parameters $\mu_a$, $\nu_a$ satisfy the balancing condition:
\be\label{bal}
\sum_{a=1}^4(\nu_a+\mu_a)=2(\omega_1+\omega_2)\equiv 2Q\, ,
\ee
and $\gamma^{(2)}(x;\omega_1,\omega_2)$ is hyperbolic gamma function, also sometimes called quantum dilogarithm. For definition of the  hyperbolic gamma function $\gamma^{(2)}(x;\omega_1,\omega_2)$ see appendix A.

It was shown in \cite{Teschner:2012em,Apresyan:2022erh} that the function \eqref{hyj} satisfies the identity:
\bea\label{newsymka}
J_h(\underline{\mu},\underline{\nu})=\prod_{j,k=1}^2\gamma^{(2)}(\nu_j+\mu_{k+2};\omega_1,\omega_2)
\prod_{j,k=1}^2\gamma^{(2)}(\nu_{j+2}+\mu_k;\omega_1,\omega_2)J_h(\underline{\tilde{\mu}},\underline{\tilde{\nu}}) \, ,
\eea
where
$$
\tilde{\mu}_1=G-\nu_1,\quad \tilde{\mu}_2=G-\nu_2,\quad \tilde{\mu}_3=Q-\nu_3,\quad \tilde{\mu}_4=Q-\nu_4,
$$
$$
\tilde{\nu}_1=-\mu_1,\quad \tilde{\nu}_2=-\mu_2,\quad \tilde{\nu}_3=Q-G-\mu_3,\quad \tilde{\nu}_4=Q-G-\mu_4,
$$
and $G=\nu_1+\nu_2+\mu_1+\mu_2$.

Now consider the limit:
\be\label{bki}
\sqrt{\omega_1\over \omega_2}=\textup{i}+\delta, \quad \delta\to 0\, .
\ee
In this limit one has as well
\be\label{qlim}
\omega_1=i\sqrt{\omega_1\omega_2}(1-i\delta),\; \omega_2=i\sqrt{\omega_1\omega_2}(-1-i\delta), \; Q=\omega_1+\omega_2=-2i\delta(i\sqrt{\omega_1\omega_2}),\; \sqrt{\omega_2\over \omega_1}=-\textup{i}+\delta.
\ee
It was conjectured in \cite{BMS} and rigorously proved in \cite{Sarkissian:2020ipg}, that in the limit \eqref{bki} the following asymptotic estimate holds uniformly on compacta:
\begin{equation}\label{gam2lim2}
\gamma^{(2)}(\textup{i}\sqrt{\omega_1\omega_2}(n+x\delta);\omega_1,\omega_2)\stackreb{=}{\delta\to 0^+} e^{\frac{\pi \textup{i}}{2}n^2} (4\pi\delta)^{\textup{i}x-1}{\bf \Gamma}(x,n),
\end{equation}
where $n\in {\mathbb Z}, \, x\in {\mathbb C}$.

We apply the limit \eqref{bki} to identity \eqref{newsymka} taking the parameters $\mu_a$ and $\nu_a$ in the following parametrization
\be\label{zbi}
\mu_a=\textup{i}\sqrt{\omega_1\omega_2}(n_a+s_a\delta),
\quad\nu_a=\textup{i}\sqrt{\omega_1\omega_2}(m_a+t_a\delta),\quad a=1,2,3,4\, ,
\ee
where $s_a, t_a\in\mathbb{C}$ and $n_a, m_a \in\Z$.
Denote the integration variables in the left and right hand side integrals in \eqref{newsymka} by $z_1$ and $z_2$ correspondingly.
We parameterize them in the same manner:
\be\label{zbi}
z_{1}=\textup{i}\sqrt{\omega_1\omega_2}(L_{1}+y_{1}\delta),\quad z_{2}=\textup{i}\sqrt{\omega_1\omega_2}(L_{2}+y_{2}\delta)\, ,
\ee
where $L_1,L_2\in \mathbb{Z}$ and $y_1,y_2\in \mathbb{C}$.
The balancing condition (\ref{bal}) together with the relations (\ref{qlim}) imply that the parameters
$n_a$, $m_a$, $s_a$ and $t_a$ satisfy the relation \eqref{baldis}.

Taking into account that in this parametrization
$$
G=i\sqrt{\omega_1\omega_2}(K+Y\delta)\, ,
$$
where
$$
K=n_1+n_2+m_1+m_2,\quad\quad Y=s_1+s_2+t_1+t_2\, ,
$$
we obtain the following transformation property of the complex hypergeometric function:
\bea\label{ide1ii}
\mathcal{J}_{cr}(\underline{s},\underline{n};\underline{t},\underline{m})=e^{\pi\textup{i}K}\prod_{ j, k =1}^2{\bf \Gamma}(t_j+s_{k+2},m_j+n_{k+2})\prod_{ j, k =1}^2{\bf \Gamma}(t_{j+2}+s_k,m_{j+2}+n_k)
\mathcal{J}_{cr}(\underline{\tilde{s}},\underline{\tilde{n}};\underline{\tilde{t}},\underline{\tilde{m}}).\quad
\end{eqnarray}
where
\bea
&&\tilde{s}_1=Y-t_1,\, \tilde{n}_1=K-m_1,\,\tilde{s}_2=Y-t_2,\, \tilde{n}_2=K-m_2,\,\tilde{s}_3=-2\imath-t_3,\, \tilde{n}_3=-m_3,\\ \nonumber
&&\tilde{s}_4=-2\imath-t_4,\, \tilde{n}_4=-m_4,\, \tilde{t}_1=-s_1,\, \tilde{m}_1=-n_1,\,\tilde{t}_2=-s_2,\, \tilde{m}_2=-n_2,\\ \nonumber
&&\tilde{t}_3=-2i-Y-s_3,\, \tilde{m}_3=-K-n_3,\,\tilde{t}_4=-2i-Y-s_4,\, \tilde{m}_4=-K-n_4,\, .
\eea

The function $\mathcal{J}_{cr}(\underline{s},\underline{n};\underline{t},\underline{m})$ here is defined in \eqref{jdiscret}.
It is easy to see that new parameters satisfy the balancing condition \eqref{baldis} as well:
\be\label{baldis2}
\sum_{a=1}^4( \tilde{n}_a+\tilde{m}_a)=0,\qquad \sum_{a=1}^4( \tilde{s}_a+\tilde{t}_a)=-4\textup{i}.
\ee
Note that in this limit contour integral in \eqref{hyj} becomes the combination of sum and integral in \eqref{jdiscret}.
This phenomenon was observed already in \cite{BMS}. It is explained there from the physical considerations, namely in this limit
the integration variables $z_1$ and $z_2$ acquire additional discrete ``degrees of freedom" $L_1$, $L_2$, and the corresponding integrals,
being from the physical point of view partition functions, should be summed up by all the state parameters.
Rigorously this phenomenon was justified in \cite{Sarkissian:2020ipg}. For readers convenience, the corresponding arguments are outlined in appendix B. Another important point here that thanks to the balancing conditions \eqref{baldis} and \eqref{baldis2}, parameter $\delta$ drops in the final limiting identity \eqref{ide1ii}.
Details of calculation can be found in appendix C.

Now let us apply \eqref{ide1ii} to the integral ${\mathcal J}_{cr}(\underline{R},\underline{S};
\underline{U},\underline{T})$ in the expression \eqref{idsgamma} setting:
\begin{equation}\label{sr1}
s_1=R_3\, ,\quad
s_2=R_1\, ,\quad
s_3=R_4\, ,\quad
s_4=R_2\, ,
\end{equation}
\begin{equation}\label{tu1}
t_1=U_3\, ,\quad
t_2=U_2\, ,\quad
t_3=U_1\, ,\quad
t_4=U_4\, ,
\end{equation}
\begin{equation}\label{ns1}
n_1=S_3\, ,\quad
n_2=S_1\, ,\quad
n_3=S_4\, ,\quad
n_4=S_2\, ,
\end{equation}
\begin{equation}\label{mt1}
m_1=T_3\, , \quad m_2=T_2\, , \quad m_3=T_1\, , \quad m_4=T_4\, .
\end{equation}
With this setting $Y$ and $K$ take the form
\be\label{ysig}
Y=-\sigma_3-\sigma_1+\rho_1-\rho_2-2\textup{i}\, ,
\ee
\be\label{kn}
K=(-N_3-N_1+M_1-M_2)/2\, .
\ee
Inserting \eqref{sr1}-\eqref{kn} in (\ref{ide1ii}) and afterwards shifting $y_2\to y_2-U_1-2i$ and $L_2\to L_2-T_1$ (remember that $T_1\in \mathbb{Z}$) we receive
\be\label{jcrn}
{\mathcal J}_{cr}(\underline{R},\underline{S};
\underline{U},\underline{T})=e^{\pi\textup{i}K}\Omega_1(\underline{\sigma},\underline{\rho},\underline{N},\underline{M})
{\mathcal J}_{cr}(\underline{\tilde{R}},\underline{\tilde{S}};
\underline{\tilde{U}},\underline{\tilde{T}})\, ,
\ee

where

\begin{equation}
\begin{aligned}
&\tilde{R}_1=\sigma_4-\sigma_2-\sigma_1-\sigma_3,\\
&\tilde{R}_2=-\rho_2-\sigma_1-\sigma_3-\rho_1,\\
&\tilde{R}_3=0,\\
&\tilde{R}_4=\sigma_4-\rho_2-\rho_1-\sigma_2,
\end{aligned}\quad
\begin{aligned}
&\tilde{U}_1=\sigma_3+\sigma_2+\rho_1-i,\\
&\tilde{U}_2=\sigma_1-\sigma_4+\rho_1-i,\\
&\tilde{U}_3=\sigma_1+\sigma_2+\rho_2-i,\\
&\tilde{U}_4=\sigma_3-\sigma_4+\rho_2-i,
\end{aligned}\quad
\begin{aligned}
&\tilde{S}_1=(N_4-N_2-N_1-N_3)/2,\\
&\tilde{S}_2=(-M_2-N_1-N_3-M_1)/2,\\
&\tilde{S}_3=0,\\
&\tilde{S}_4=(N_4-M_2-M_1-N_2)/2,
\end{aligned}
\end{equation}
\begin{equation}
\tilde{T}_1=(N_3+N_2+M_1)/2, \quad \tilde{T}_2=(N_1-N_4+M_1)/2, \quad \tilde{T}_3=(N_1+N_2+M_2)/2, \quad \tilde{T}_4=(N_3-N_4+M_2)/2,
\end{equation}
and
\bea
&&\Omega_1(\underline{\sigma},\underline{\rho},\underline{N},\underline{M})=\\ \nonumber
&&{\bf \Gamma}(-\sigma_3-\sigma_4+\rho_2-\textup{i},(-N_3-N_4+M_2)/2)
{\bf \Gamma}(-\sigma_1+\sigma_2+\rho_2-\textup{i},(-N_1+N_2+M_2)/2)\qquad\\ \nonumber
&&{\bf \Gamma}(-\rho_1-\sigma_2-\sigma_3-\textup{i},-(M_1-N_2-N_3)/2){\bf \Gamma}(-\rho_1+\sigma_4-\sigma_1-\textup{i},(-M_1+N_4-N_1)/2)\qquad\\ \nonumber
&&{\bf \Gamma}(\sigma_3-\sigma_2+\rho_1-\textup{i},(N_3-N_2+M_1)/2){\bf \Gamma}(\sigma_3-\sigma_4-\rho_2-\textup{i},(N_3-N_4-M_2)/2)\qquad\\ \nonumber
&&{\bf \Gamma}(\sigma_4+\sigma_1+\rho_1-\textup{i},(N_4+N_1+M_1)/2){\bf \Gamma}(\sigma_2+\sigma_1-\rho_2-\textup{i},(N_2+N_1-M_2)/2)\, .
\eea

Inserting \eqref{jcrn} in \eqref{idsgamma} we get new expression for $6j$-symbols of the $SL(2,\mathbb{C})$ group:
\bea \label{idsgammaa}
\big\{\,{}^{\sigma_1,N_1}_{\sigma_3,N_3}\,{}^{\sigma_2,N_2}_{{\sigma}_4,N_4}\,|\,{}^{\rho_1,M_1}_{\rho_2,M_2}\big\}
=(-1)^{M_2-N_2+N_4}e^{\pi\textup{i}K}{\pi^2\over 4}\Omega_2(\underline{\sigma},\underline{\rho},\underline{N},\underline{M}){\mathcal J}_{cr}(\underline{\tilde{R}},\underline{\tilde{S}};
\underline{\tilde{U}},\underline{\tilde{T}})\, ,
\eea
where
\bea
&&\Omega_2(\underline{\sigma},\underline{\rho},\underline{N},\underline{M})=\\ \nonumber
&&{\bf \Gamma}(\sigma_1-\sigma_2+\rho_2-\textup{i},(N_1-N_2+M_2)/2)
{\bf \Gamma}(-\sigma_1+\sigma_2+\rho_2-\textup{i},(-N_1+N_2+M_2)/2)\qquad\\ \nonumber
&&{\bf \Gamma}(-\rho_1-\sigma_2-\sigma_3-\textup{i},-(M_1-N_2-N_3)/2){\bf \Gamma}(-\rho_1+\sigma_4-\sigma_1-\textup{i},(-M_1+N_4-N_1)/2)\qquad\\ \nonumber
&&{\bf \Gamma}(\sigma_3-\sigma_2+\rho_1-\textup{i},(N_3-N_2+M_1)/2){\bf \Gamma}(\sigma_3-\sigma_4-\rho_2-\textup{i},(N_3-N_4-M_2)/2)\qquad\\ \nonumber
&&{\bf \Gamma}(\sigma_2-\sigma_3+\rho_1-\textup{i},(N_2-N_3+M_1)/2){\bf \Gamma}(\sigma_2+\sigma_1-\rho_2-\textup{i},(N_2+N_1-M_2)/2)\, .
\eea
Define
\bea\nonumber
&&{\mathcal S}={1\over 2}(\sigma_1+\sigma_2+\sigma_3-\sigma_4)\, ,\\ \nonumber
&&{\mathcal N}={1\over 2}(N_1+N_2+N_3-N_4)\, .\\ \nonumber
\eea
Since $A_1$ and $A_3$ are integer, the equality $A_1-A_3={1\over 2}(N_1-N_2+N_3+N_4)$ implies that $N_1-N_2+N_3+N_4$ is even and  ${\mathcal N}$ is integer,
and we are allowed to define the following analogue of the Regge transformation:
\bea\label{transfor}
&&\sigma_1\to {\mathcal S}-\sigma_1,\quad \sigma_2\to {\mathcal S}-\sigma_2,\quad \sigma_3\to {\mathcal S}-\sigma_3,\quad \sigma_4\to -({\mathcal S}+\sigma_4)\, ,\\ \nonumber
&&N_1\to {\mathcal N}-N_1,\; N_2\to {\mathcal N}-N_2,\; N_3\to {\mathcal N}-N_3,\;  N_4\to -({\mathcal N}+N_4)\, .
\eea
It is easy to check that the transformation \eqref{transfor} leaves intact the $\Omega_2(\underline{\sigma},\underline{\rho},\underline{N},\underline{M})$
and ${\mathcal J}_{cr}(\underline{\tilde{R}},\underline{\tilde{S}};
\underline{\tilde{U}},\underline{\tilde{T}})$ just reshuffling the entering there terms and it remains only to check
the transformation of the sign prefactor.
In particular one has that under the transformation \eqref{transfor}:
\be
N_1+N_3\to N_2-N_4\quad {\rm and}\quad -N_2+N_4\to -N_1-N_3\, .
\ee
Since $-N_2+N_4+N_1+N_3$ is even, the first factor is invariant under \eqref{transfor}, and the second factor changes
by sign $(-1)^{(N_1+N_3-N_2+N_4)/2}$. Therefore we obtain that under \eqref{transfor} the $6j$-symbols \eqref{idsgamma}
may acquire only a sign:

\bea \nonumber
\big\{\,{}^{{\mathcal S}-\sigma_1,{\mathcal N}-N_1}_{{\mathcal S}-\sigma_3,{\mathcal N}-N_3}\,{}^{{\mathcal S}-\sigma_2,{\mathcal N}-N_2}_{{-({\mathcal S}+\sigma}_4),-({\mathcal N}+N_4)}\,|\,{}^{\rho_1,M_1}_{\rho_2,M_2}\big\}
=
(-1)^{(N_1+N_3-N_2+N_4)/2}\big\{\,{}^{\sigma_1,N_1}_{\sigma_3,N_3}\,{}^{\sigma_2,N_2}_{{\sigma}_4,N_4}\,|\,{}^{\rho_1,M_1}_{\rho_2,M_2}\big\}\, .
\eea
Thus we have established that the $6j$-symbols for the unitary principal series representations of the $SL(2,\mathbb{C})$ group are
invariant under the action of the generalization of the Regge transformation \eqref{transfor} up to the possible sign factor.

\section{Conclusion}
In this paper we obtain new symmetry transformation \eqref{ide1ii} of the complex hypergeometric function \eqref{jdiscret},
which is essential part of the $6j$-symbols for the unitary principal series representations of the $SL(2,\mathbb{C})$ group.
This allowed us to obtain new expression \eqref{idsgammaa} for them. This expression makes obvious presence of the Regge
symmetry \eqref{transfor}.
At the next step we are planning using the expression \eqref{idsgammaa} to study asymptotic form of $6j$-symbols
for the large spins. We hope that it will uncover complex non-compact counterpart of the Ponzano-Regge asymptotic formulae.

\vspace{0.5cm}

{\bf Acknowledgements.}
The work of Elena Apresyan was supported by  Armenian SCS grants 21AG‐1C024 and 24FP-1F039. We would like also to thank
V. P. Spiridonov for many useful discussions.

\appendix

\section{Hyperbolic gamma function}
The function $\gamma^{(2)}(y;\omega_1,\omega_2)$ has the integral representation \cite{Spiridonov:2010em}
\be
\gamma^{(2)}(y;\omega_1,\omega_2)=\exp\left(-\int_0^{\infty}\left({\sinh(2y-\omega_1-\omega_2)x\over 2\sinh(\omega_1x)
\sinh(\omega_2x)}-{2y-\omega_1-\omega_2\over 2\omega_1\omega_2x}\right)\right){dx\over x}\, ,
\ee
and obeys the equations:
\be\label{hp1}
{\gamma^{(2)}(y+\omega_1;\omega_1,\omega_2)\over \gamma^{(2)}(y;\omega_1,\omega_2)}=2\sin{\pi y\over \omega_2}\,  ,\quad
{\gamma^{(2)}(y+\omega_2;\omega_1,\omega_2)\over \gamma^{(2)}(y;\omega_1,\omega_2)}=2\sin{\pi y\over \omega_1}.
\ee
\section{Limit of integrals}

Here we show that in the limit \eqref{bki} the integral of a product of hyperbolic gamma functions yields the infinite sum
of integrals of the limit of the integrand \cite{Sarkissian:2020ipg}.

Consider the integral:
$$
\int_{-\textup{i}\infty}^{\textup{i}\infty}\Delta(z){dz\over \textup{i}\sqrt{\omega_1\omega_2}}
=\int_{-\textup{i}\infty}^{\textup{i}\infty}\Delta(\sqrt{\omega_1\omega_2}x){dx\over \textup{i}},
\quad x=\frac{z}{\sqrt{\omega_1\omega_2}},
$$
where $\Delta(z)$ is a product of $\gamma^{(2)}(u;\omega_1,\omega_2)$.
We can rewrite it as an infinite sum
$$
\int_{-\textup{i}\infty}^{\textup{i}\infty}\Delta(\sqrt{\omega_1\omega_2}\, x){dx\over \textup{i}}=
\sum_{N\in \mathbb{Z}} \int_{\textup{i}(N-1/2)}^{\textup{i}(N+1/2)}\Delta(\sqrt{\omega_1\omega_2}\, x){dx\over \textup{i}}
$$
$$
= \sum_{N\in \mathbb{Z}} \int_{N-1/2}^{N+1/2}\Delta( \textup{i}\sqrt{\omega_1\omega_2}\, x)dx
=\sum_{N\in \mathbb{Z}} \int_{-1/2}^{1/2} \Delta(\textup{i}\sqrt{\omega_1\omega_2}(N+x))dx.
$$
At the second step we changed the variable $x\to \imath x$ and at the last step performed shift $x\to x+N$.
Now we parametrise $x=y\delta, \, \delta>0,$ and take the limit $\delta\to 0^+$:
\be  \makebox[-6em]{}
{\rm lim}_{\delta\to 0}\sum_{N\in \mathbb{Z}} \int_{-1/2}^{1/2}
\Delta(\textup{i}\sqrt{\omega_1\omega_2}(N+x))dx
= {\rm lim}_{\delta\to 0}
\sum_{N\in \mathbb{Z}} \int_{-1/2\delta}^{1/2\delta}
\delta \Delta(\textup{i}\sqrt{\omega_1\omega_2}(N+y\delta))dy.
\ee
The sum over $N$ is infinite, for $\delta\to 0^+$ the integration contour becomes real axis $(-\infty,\infty)$.
Using the uniform convergence of the limit \eqref{gam2lim2} we interchange the limit  ${\rm lim}_{\delta\to 0}$ with
the summation and integration and end up with
\be\label{limintigr}
\sum_{N\in \mathbb{Z}}
\int_{-\infty}^{\infty} \left[
{\rm lim}_{\delta\to 0}\delta\Delta(\textup{i}\sqrt{\omega_1\omega_2}(N+y\delta))\right] dy.
\ee

\section{Details of calculation of identity \eqref{ide1ii}}

Here we give the details of calculation of identity \eqref{ide1ii}.
In this section we use notation $\gamma^{(2)}(y;\mathbf{\omega})\equiv\gamma^{(2)}(y;\omega_1,\omega_2)$.
At the beginning let us present the asymptotic forms in the limit $\delta\to 0$ for the hyperbolic gamma functions
used in formula \eqref{newsymka}:
\bea\label{limgamfor}
&&\gamma^{(2)}(\mu_a-z_1;\mathbf{\omega})\to e^{\frac{\pi \textup{i}}{2}(n_a-L_1)^2} (4\pi\delta)^{\textup{i}(s_a-y_1)-1}{\bf \Gamma}(s_a-y_1,n_a-L_1)\quad a=1,2,3,4\, ,\\ \nonumber
&&\gamma^{(2)}(\nu_a+z_1;\mathbf{\omega})\to e^{\frac{\pi \textup{i}}{2}(m_a+L_1)^2} (4\pi\delta)^{\textup{i}(t_a+y_1)-1}{\bf \Gamma}(t_a+y_1,m_a+L_1)\quad a=1,2,3,4\, ,\\ \nonumber
&&\gamma^{(2)}(\nu_j+\mu_{k+2};\mathbf{\omega})\to e^{\frac{\pi \textup{i}}{2}(m_j+n_{k+2})^2} (4\pi\delta)^{\textup{i}(t_j+s_{k+2})-1}{\bf\Gamma}(t_j+s_{k+2},m_j+n_{k+2})\; j,k=1,2\, ,\\ \nonumber
&&\gamma^{(2)}(\nu_{j+2}+\mu_{k};\mathbf{\omega})\to e^{\frac{\pi \textup{i}}{2}(m_{j+2}+n_{k})^2} (4\pi\delta)^{\textup{i}(t_{j+2}+s_{k})-1}{\bf\Gamma}(t_{j+2}+s_{k},m_{j+2}+n_{k})\; j,k=1,2\, ,\\ \nonumber
&&\gamma^{(2)}(G-\nu_1-z_2;\mathbf{\omega})\to e^{\frac{\pi \textup{i}}{2}(K-m_1-L_2)^2} (4\pi\delta)^{\textup{i}(Y-t_1-y_2)-1}{\bf \Gamma}(Y-t_1-y_2,K-m_1-L_2)\, ,\\ \nonumber
&&\gamma^{(2)}(G-\nu_2-z_2;\mathbf{\omega})\to e^{\frac{\pi \textup{i}}{2}(K-m_2-L_2)^2} (4\pi\delta)^{\textup{i}(Y-t_2-y_2)-1}{\bf \Gamma}(Y-t_2-y_2,K-m_2-L_2)\, ,\\ \nonumber
&&\gamma^{(2)}(Q-\nu_3-z_2;\mathbf{\omega})\to e^{\frac{\pi \textup{i}}{2}(-m_3-L_2)^2} (4\pi\delta)^{\textup{i}(-2i-t_3-y_2)-1}{\bf \Gamma}(-2i-t_3-y_2,-m_3-L_2)\, ,\\ \nonumber
&&\gamma^{(2)}(Q-\nu_4-z_2;\mathbf{\omega})\to e^{\frac{\pi \textup{i}}{2}(-m_4-L_2)^2} (4\pi\delta)^{\textup{i}(-2i-t_4-y_2)-1}{\bf \Gamma}(-2i-t_4-y_2,-m_4-L_2)\, ,\\ \nonumber
&&\gamma^{(2)}(-\mu_1+z_2;\mathbf{\omega})\to e^{\frac{\pi \textup{i}}{2}(-n_1+L_2)^2} (4\pi\delta)^{\textup{i}(-s_1+y_2)-1}{\bf \Gamma}(-s_1+y_2,-n_1+L_2)\, ,\\ \nonumber
&&\gamma^{(2)}(-\mu_2+z_2;\mathbf{\omega})\to e^{\frac{\pi \textup{i}}{2}(-n_2+L_2)^2} (4\pi\delta)^{\textup{i}(-s_2+y_2)-1}{\bf \Gamma}(-s_2+y_2,-n_2+L_2)\, ,\\ \nonumber
&&\gamma^{(2)}(Q-G-\mu_3+z_2;\mathbf{\omega})\to e^{\frac{\pi \textup{i}}{2}(-K-n_3+L_2)^2} (4\pi\delta)^{\textup{i}(-2i-Y-s_3+y_2)-1}{\bf \Gamma}(-2i-Y-s_3+y_2,-K-n_3+L_2)\, ,\\ \nonumber
&&\gamma^{(2)}(Q-G-\mu_4+z_2;\mathbf{\omega})\to e^{\frac{\pi \textup{i}}{2}(-K-n_4+L_2)^2} (4\pi\delta)^{\textup{i}(-2i-Y-s_4+y_2)-1}{\bf \Gamma}(-2i-Y-s_4+y_2,-K-n_4+L_2)\, .
\eea
Using first two lines in \eqref{limgamfor} and formula \eqref{limintigr}, we obtain :
\bea
&&J_h(\mu,\nu)\to  e^{\frac{\pi \textup{i}}{2}\sum_{a=1}^4(n_a^2+m_a^2)} (4\pi\delta)^{\textup{i}\sum_{a=1}^4(t_a+s_a)-8}(4\pi\delta)\mathcal{J}_{cr}(\underline{s},\underline{n};\underline{t},\underline{m})\\ \nonumber
&&=e^{\frac{\pi \textup{i}}{2}\sum_{a=1}^4(n_a^2+m_a^2)} (4\pi\delta)^{-3}\mathcal{J}_{cr}(\underline{s},\underline{n};\underline{t},\underline{m})
\eea
The terms with $L_1$ in the phase factors drop  being multiple of $e^{2\pi \imath}$ due to the discrete part of the balancing condition  \eqref{baldis}. At the second step we used also the balancing condition  \eqref{baldis} for the continuous part.
Similarly one obtains using the last eight lines in \eqref{limgamfor}, formula \eqref{limintigr} and the balancing condition \eqref{baldis2}:
\bea
&&J_h(\tilde{\mu},\tilde{\nu})\to  e^{\frac{\pi \textup{i}}{2}\sum_{a=1}^4(\tilde{n}_a^2+\tilde{m}_a^2)} (4\pi\delta)^{\textup{i}\sum_{a=1}^4(\tilde{t}_a+\tilde{s}_a)-8}(4\pi\delta)\mathcal{J}_{cr}(\underline{\tilde{s}},\underline{\tilde{n}};
\underline{\tilde{t}},\underline{\tilde{m}})\\ \nonumber
&&=e^{\frac{\pi \textup{i}}{2}\sum_{a=1}^4(\tilde{n}_a^2+\tilde{m}_a^2)} (4\pi\delta)^{-3}\mathcal{J}_{cr}(\underline{\tilde{s}},\underline{\tilde{n}};\underline{\tilde{t}},\underline{\tilde{m}})
\eea
And finally using third and fourth line we obtain:
\bea
&&\prod_{j,k=1}^2\gamma^{(2)}(\nu_j+\mu_{k+2};\mathbf{\omega})\prod_{j,k=1}^2\gamma^{(2)}(\nu_{j+2}+\mu_k;\mathbf{\omega})
\to e^{\frac{\pi \textup{i}}{2}\sum_{ j, k =1}^2((m_j+n_{k+2})^2+(m_{j+2}+n_{k})^2)} \nonumber
\\
&&\times (4\pi\delta)^{2\textup{i}\sum_{a=1}^4(t_a+s_a)-8}\prod_{ j, k =1}^2{\bf \Gamma}(t_j+s_{k+2},m_j+n_{k+2})\prod_{ j, k =1}^2{\bf \Gamma}(t_{j+2}+s_k,m_{j+2}+n_k)\\
&&=e^{\frac{\pi \textup{i}}{2}\sum_{ j, k =1}^2((m_j+n_{k+2})^2+(m_{j+2}+n_{k})^2)} \prod_{ j, k =1}^2{\bf \Gamma}(t_j+s_{k+2},m_j+n_{k+2})\prod_{ j, k =1}^2{\bf \Gamma}(t_{j+2}+s_k,m_{j+2}+n_k)\qquad
\eea
At the second step we used again the balancing condition  \eqref{baldis}.
Collecting all we see that the factor $\delta$ drops everywhere, and simplifying phase factors we derive \eqref{ide1ii}.

\end{document}